\documentclass[aip,amsmath,amssymb,reprint]{revtex4-1}
\usepackage{graphicx}
\usepackage{dcolumn}
\usepackage{bm}
\usepackage{color}

\usepackage[utf8]{inputenc}
\usepackage[T1]{fontenc}
\usepackage{mathptmx}
\usepackage{etoolbox}

\makeatletter
\def\@email#1#2{%
  \endgroup
  \patchcmd{\titleblock@produce}
   {\frontmatter@RRAPformat}
   {\frontmatter@RRAPformat{\produce@RRAP{*#1\href{mailto:#2}{#2}}}\frontmatter@RRAPformat}
   {}{}
}%
\makeatother
\begin{document}

\preprint{AIP/123-QED}

\title[Driven Probe Particle Dynamics in a Bubble Forming System]{Driven Probe Particle Dynamics in a Bubble Forming System}
\author{
C. Reichhardt and C. J. O. Reichhardt 
}
\email{cjrx@lanl.gov}
\affiliation{
Theoretical Division and Center for Nonlinear Studies,
Los Alamos National Laboratory, Los Alamos, New Mexico 87545, USA
}%

\date{\today}

\begin{abstract}
We numerically examine the dynamics of a probe particle driven at a constant force through an assembly of particles with competing long-range repulsion and short-range attraction that forms a bubble or stripe state. In the bubble regime, we identify several distinct types of motion, including an elastic or pinned regime where the probe particle remains inside a bubble and drags all other bubbles with it. There is also a plastic bubble phase where the bubble in which the probe particle is trapped is able to move past the adjacent bubbles. At larger drives, there is a breakthrough regime where the probe particle jumps from bubble to bubble, and in some cases, can induce correlated rotations or plastic rearrangements of the particles within the bubbles. At the highest drives, the probe particle moves sufficiently rapidly that the background particles undergo only small distortions. The distinctive dynamic flow states and the transitions between them are accompanied by signatures in the effective drag on the driven particle, jumps in the velocity-force curves, and changes in the time-dependent velocity fluctuations. We map the dynamic phase diagram for this system for varied interaction lengths, bubble sizes, and densities.
\end{abstract}
\maketitle

\section{Introduction}

The local dynamic response of a system can be examined by
measuring the velocity-force, drag, or fluctuation response of a
single driven particle moving through the medium
\cite{Hastings03,Habdas04,Reichhardt04a,Squires05,Gazuz09,Khair10,Winter12,Zia18}.
This method is known as active microrheology,
and it can be used to detect
changes that occur in the threshold force
needed to drive the particle through the medium
as the density of the system is changed or as a phase transition
occurs
\cite{Hastings03,Habdas04,Benichou13a,Puertas14,Senbil19,Gruber20,Reichhardt22b}.
The response of the probe particle will depend on whether the medium is a crystal \cite{Reichhardt04a}, amorphous solid, or fluid \cite{Habdas04}, and
will differ when the medium is unable to respond to the rapid motion
of the probe particle
compared to when the probe particle generates plastic deformations
in the medium
\cite{Hastings03}.
Active microrheology has been studied in a variety of soft matter systems,
including
colloidal assemblies \cite{Habdas04,Zia18},
granular matter \cite{Drocco05,Candelier10,Kolb13},
and active matter \cite{Reichhardt15,Knezevic21,Peng22}.
Similar methods have also been applied to hard condensed matter systems such as superconductors \cite{Straver08,Auslaender09,Reichhardt09a}
and magnetic systems,
where a tip can drag a superconducting vortex or a skyrmion
\cite{Wang20b,Reichhardt21a}.

In most systems where active microrheology has been studied,
the background particles have either a short-range repulsion,
as in granular matter or hard-sphere colloids,
or a softer intermediate-range repulsion, as in charged colloids,
superconducting vortices, and magnetic skyrmions.
Active microrheology can also be applied to systems with competing
short-range attractive and long-range repulsive (SALR) interaction
potentials \cite{Hooshanginejad24}, where complex structures can
emerge
such as modulated bubbles of various sizes, stripes,
and void crystals \cite{Seul95,Stoycheva00,Reichhardt03,Reichhardt04,Sciortino04,Nelissen05,Liu08,Liu19,Hooshanginejad24,Apolinario25}.
When the interaction length in a SALR system is held fixed while
the particle density is increased, the system successively passes through
dilute crystal, bubble, stripe, void lattice, and dense crystal states
\cite{Seul95,Stoycheva00,Reichhardt03,Reichhardt04,Mossa04,Sciortino04,Nelissen05,Liu08,Reichhardt10,McDermott16,Liu19,Hooshanginejad24}.
Additional sub-phases can be identified, such as stripes that are
either one particle or multiple particles wide, or bubbles that are
as small as two particles per bubble or composed of a large number of
particles.
Similar patterns can form in systems where the interactions are purely
repulsive when there are multiple length scales or steps in the
repulsive potential
\cite{Jagla98,Malescio03,Glaser07}.

In soft matter systems, stripe and bubble phases can
occur in colloidal assemblies, emulsions, binary fluids, and
magnetic colloids \cite{Malescio03,Glaser07,CostaCampos13}, while 
in hard condensed matter systems, they can appear
for electron liquid crystal states in magnetic fields
\cite{Fogler96,Moessner96,Cooper99,Pan99,Fradkin99,Gores07,Zhu09,Friess18,Shingla23}. 
In some charge-ordered hard matter systems, a competition between the
Coulomb repulsion and short-range attraction from spin or strains
generates similar patterns
\cite{Tranquada95,Reichhardt04a,Mertelj05}.
Bubbles and stripes can also form in vortex phases in multiband superconductors 
\cite{Xu11,Komendova13,Varney13,Sellin13,Brems22}
and in skyrmion-vortex hybrid systems \cite{Reichhardt22a}.

In a recent study \cite{Reichhardt25}, active rheology was
applied to a pattern forming SALR system.
At small drives, the response of the system is elastic and the entire
particle assembly is dragged along by the probe particle, but above
a critical depinning force $F_c$, the probe particle breaks away and
moves through rather than with the background particles.
For fixed particle density and increasing attraction strength,
a minimum of $F_c$ appears in the stripe state, and the threshold
force increases rapidly
in the bubble phase.
Several distinct dynamical phases were identified, including
plastic flow, in which the probe particle generates
plastic rearrangements in the surrounding medium, and a so-called
viscous flow regime, where the probe particle moves sufficiently
rapidly that no plastic rearrangements can occur because the background
particles do not have enough time to react before the probe particle has
passed by completely.
This previous work focused on the stripe and crystal phases under
conditions where the length scales of the interaction potential were
held fixed and the relative strength of the attractive and repulsive
terms was varied.

In the present work, we extend the active rheology study to systems
in which the magnitude of the interaction terms are held fixed but the
interaction length scale is varied, and we focus
on the bubble regimes for varied bubble sizes.
When the bubbles are small,
we observe three phases. There is an elastic regime in which
the probe particle remains inside a particular bubble and drags all
of the other bubbles along with it,
a plastic bubble state in which the bubble containing the probe
particle breaks away from the other bubbles and travels between them,
and a bubble-to-bubble hopping state where the probe particle is able to
jump from one bubble to another, producing strong local plastic
rearrangements in the system.
At higher drives in the bubble-to-bubble
hopping regime, we find viscous flow
in which the probe particle
moves sufficiently rapidly that plastic deformations do not
occur and individual bubbles execute small rotations
as the probe particle moves past. 
We map the evolution of these dynamic states as the interaction
length scale is swept through
crystal, stripe, and bubble phases, and find that the threshold
for plastic motion is smallest in the stripe phase. 
We also study the variation of
the dynamics for bubble sizes ranging from dimer bubbles up to
bubbles containing large numbers of particles.

\section{Simulation}

We consider a two-dimensional system
of size $L \times L$
with periodic boundary conditions in the $x$- and $y$-directions.
We place $N$ particles within the system
to obtain a system density of $\rho = N/L^2$.
The interparticle interaction potential is given by
\begin{equation}
  	V(R_{ij}) = \frac{1}{R_{ij}} - B\exp(-R_{ij}/\xi) \ .
\end{equation}
The first term represents long range repulsion,
and the second term is a phenomenological
short-range attraction with length scale $\xi$, screening length
$\kappa=1/\xi$,
and strength $B$.
Previous work on this model performed for fixed $\kappa$ and $\rho$
but varied $B$ showed that the system
forms a crystal at low $B$, stripes at intermediate $B$,
and bubbles at high $B$ \cite{Reichhardt03,Reichhardt24}.
When $\rho$ is varied at fixed $\kappa$ and large fixed $B$,
a bubble lattice appears for low $\rho$,
stripes for intermediate $\rho$,
and a uniform crystal for high $\rho$ \cite{Reichhardt10}.
In the present work,
we fix $B=5.6$ while varying $\rho$ and $\xi$.
The particle dynamics obey
an overdamped equation of motion,
and the particles are initially placed in a
lattice and allowed to relax
for a fixed number time steps until the motion ceases.
We have also tested a simulated annealing procedure in which
we start from a high temperature and gradually
cool to $T = 0.0$, and obtain similar patterned structures.
After the system has relaxed, we apply a driving force only to the
probe particle $p$. The magnitude of the driving force is increased
incrementally, and we hold the drive fixed at each increment long enough
to avoid measuring any transient effects.
The overdamped equation of motion for particle $i$ is
\begin{equation}
\eta \frac{d {\bf R}_{i}}{dt} =
-\sum^{N}_{j \neq i} \nabla V(R_{ij}) +
        {\bf F}_{D}^i ,
\end{equation}
where the driving term is applied only to particle $i=p$ and is
absent for the background particles.
Here,
the location of particle $i (j)$ is ${\bf R}_{i (j)}$,
$R_{ij}=|{\bf R}_i-{\bf R}_j|$,
and the damping term $\eta = 1.0$.
For computational efficiency, we
use a real-space Lekner summation technique to treat
the long-range interaction,
as in previous studies
\cite{Reichhardt03,Reichhardt10,McDermott14,Reichhardt24}.
We measure the average velocity of the probe particle, the time-dependent fluctuations, and the structural rearrangements.

\section{Results}

\begin{figure}
\includegraphics[width=\columnwidth]{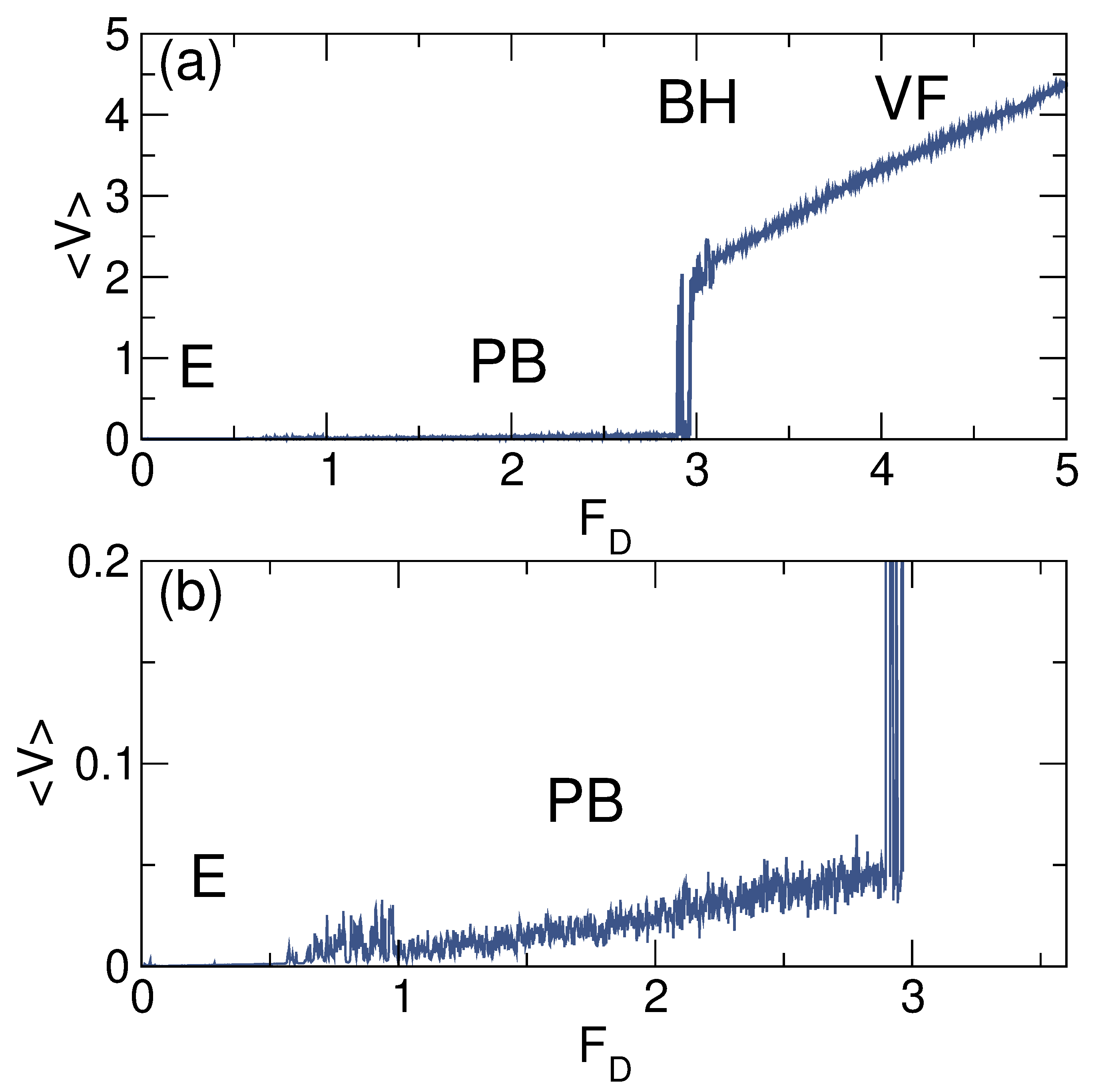}
\caption{(a) Velocity $\langle V\rangle$ vs driving force $F_D$ for a system in the bubble state at
$\xi = 0.7$ and $\rho = 0.32$.
$E$: elastic phase in which the probe particle drags all of the bubbles with it.
$PB$: plastic bubble phase in which the bubble containing the probe particle
breaks free from the other bubbles and moves between them.
$BH$: disordered bubble hopping regime.
$VF$: high-drive viscous flow regime.
(b) A blow-up of panel (a) to show more clearly the elastic phase E.
} 
\label{fig:1}
\end{figure}

\begin{figure}
\includegraphics[width=\columnwidth]{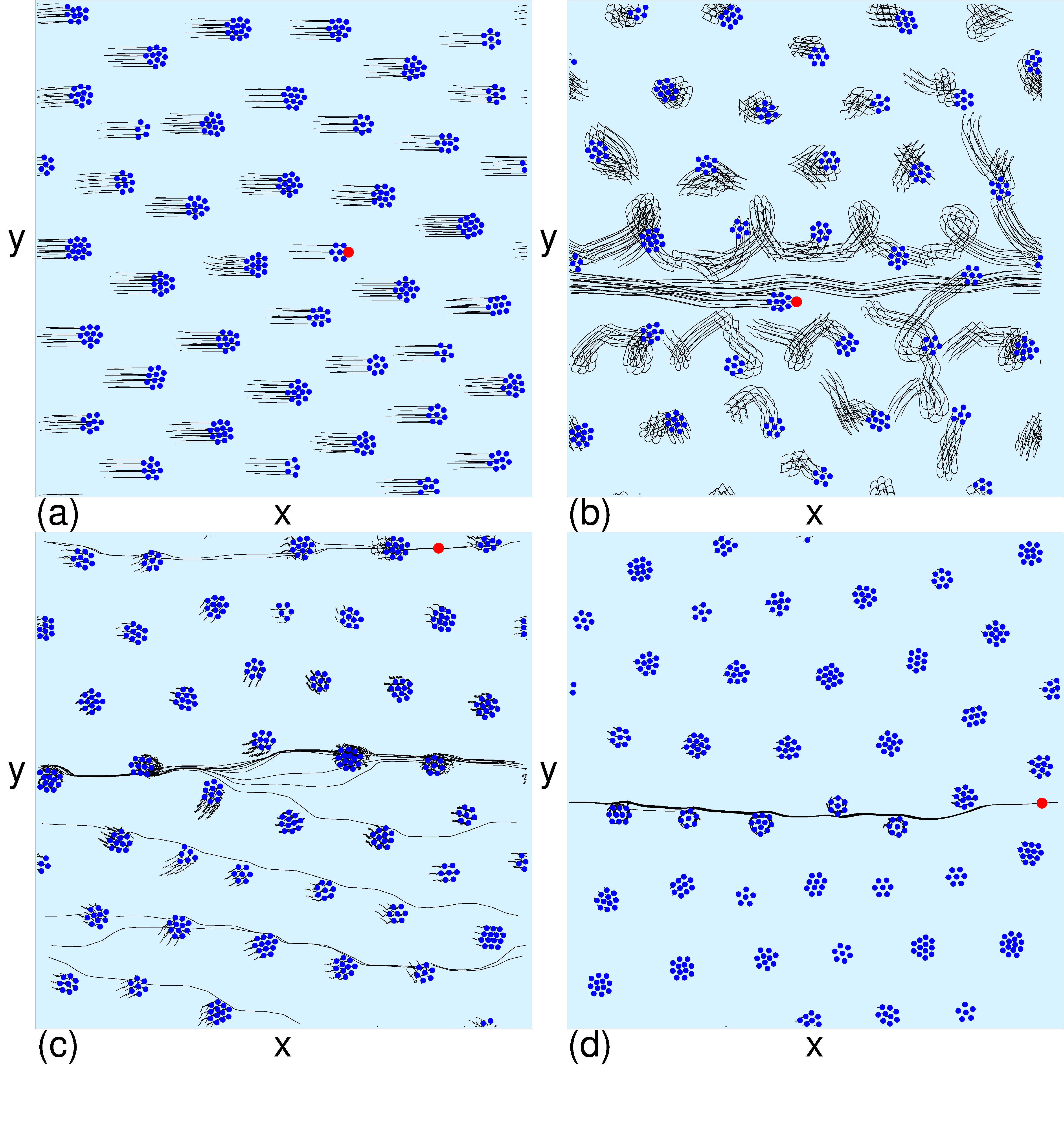}
\caption{The probe particle (red), background particles (blue), and
particle trajectories (lines) for the system from Fig.~\ref{fig:1}
in the bubble state at
$\xi=0.7$ and $\rho=0.32$.
(a) The elastic phase $E$ at
$F_D = 0.3$, where the probe particle moves all of the bubbles together.
(b) The plastic bubble flow phase $PB$ at $F_D = 1.0$,
where the bubble containing the probe particle breaks free
and moves independently of the other bubbles.
(c) The disordered bubble-to-bubble hopping regime $BH$ at $F_D = 3.0$.
(d) The viscous flow regime $VF$ at $F_D = 4.0$,
where the probe particle generates only small rotations in the bubbles
it passes.}
\label{fig:2}
\end{figure}

In Fig.~\ref{fig:1}(a), we plot the velocity $\langle V \rangle$
versus driving force $F_D$ for a system in
the bubble state at
$\xi = 0.7$ and $\rho = 0.32$.
Based on the features in the velocity-force ($v$-$f$) curve, we identify four
phases. At low drives, there is
an elastic phase $E$ where the probe particle remains trapped in
a bubble and drags all of the other bubbles along with it as it moves,
as illustrated in Fig.~\ref{fig:2}(a) at $F_D=0.3$.
A blow up of the 
$\langle V \rangle$ versus $F_D$ curve appears in Fig.~\ref{fig:1}(b)
to more clearly highlight
the elastic regime in which $\langle V \rangle = F_D/N$,
where $N$ is the number of particles in the system.
Since all of the particles move together with the probe particle
in the elastic regime, as $N$ grows, $\langle V \rangle$ decreases,
and the elastic regime can be regarded as effectively pinning the
driven particle.
For $F_D > 0.6$, the probe particle remains trapped in
a single bubble, but that bubble
breaks free from the other bubbles and the system enters
the plastic bubble flow phase, shown in Fig.~\ref{fig:2}(b) at $F_D = 1.0$.
Here the velocity
remains fairly low and is given by
$v \sim F_D/N_b$, where $N_b$ is the number of particles
in the bubble containing the probe particle.
The other bubbles generate a strong drag effect since, as seen
in Fig.~\ref{fig:2}(b), the driven bubble drags the background bubbles
a small distance before breaking away from them.
As $F_D$ increases, the driven bubble moves more rapidly
until a transition occurs to a state in which the probe particle is able
to escape from an individual bubble and begins to hop
from one bubble to another.
This transition is accompanied by
the large jump up in velocity near $F_D = 2.9$ in Fig.~\ref{fig:1}(a).
The trajectories of the particles in the bubble-to-bubble disordered
flow regime are shown at $F_D = 3.0$ in Fig.~\ref{fig:2}(c).
There are still large fluctuations in the velocity of the probe particle
since it can be trapped
temporarily in a given bubble until it can generate a sufficiently large
plastic deformation to jump out and move on to the next bubble.
For even higher drives,
the probe particle moves more rapidly than the ability of the background
particles to react,
and the probe particle only induces small rotations of the bubbles as
it passes them,
as shown in Fig.~\ref{fig:2}(d) at $F_D = 4.0$.
We call this high drive phase the viscous flow regime,
since the driven particle is moving
at nearly the velocity of a free particle, but experiences
some additional damping due to the interactions with the
background particles.
In previous work on active rheology in pattern-forming systems,
the elastic, disordered hopping, and viscous flow phases were
observed; however, the plastic bubble flow phase did not appear \cite{Reichhardt25}.

\begin{figure}
\includegraphics[width=\columnwidth]{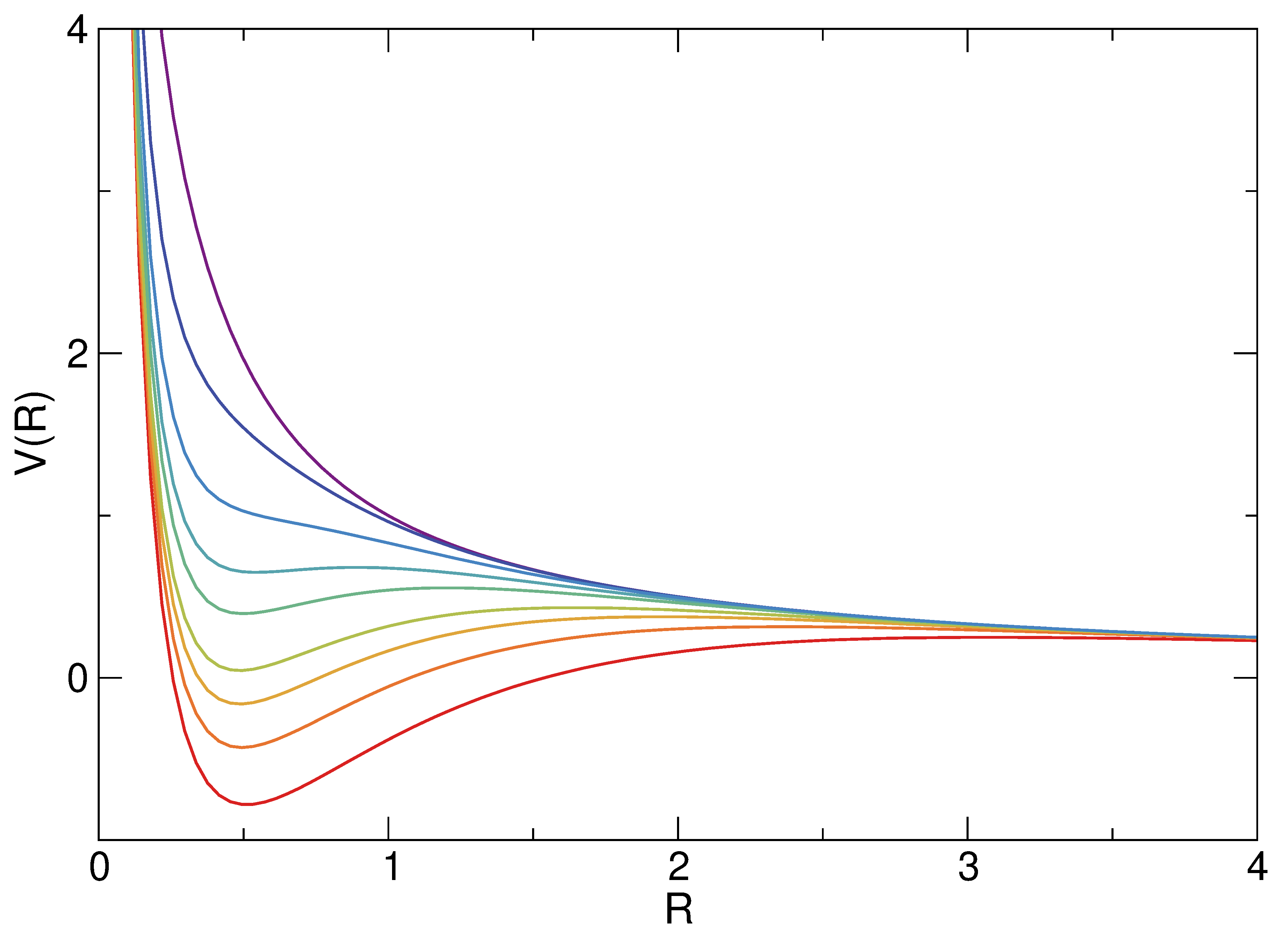}
\caption{Plot of the interaction potential $V(R)$ vs $R$
for the system in Fig.~\ref{fig:1}
with fixed $\rho=0.32$ at
$\xi = 0.7$, 0.6, 0.525, 0.475, 0.4, 0.35, 0.2, and $0.1$,
from bottom to top. 
The system forms bubbles for
$\xi > 0.55$,
stripes for $0.475 < \xi < 0.55$, and
a crystal for $\xi < 0.475$.}
\label{fig:3}
\end{figure}

We next study the evolution of the velocity-force curves as we
vary $\xi$ while fixing
$\rho=0.32$.
In Fig.~\ref{fig:3}, we plot the interaction potential
$V(R)$ versus $R$ for the system
from Fig.~\ref{fig:1}
at $\xi = 0.7$, 0.6, 0.525, 0.475, 0.4, 0.35, 0.2, and $0.1$.
As $\xi$ decreases, the repulsive term in the interaction potential
becomes more dominant, and the system forms
bubbles for $\xi > 0.55$, stripes for $0.475 < \xi < 0.55$,
and a crystal for $\xi < 0.475$.

\begin{figure}
\includegraphics[width=\columnwidth]{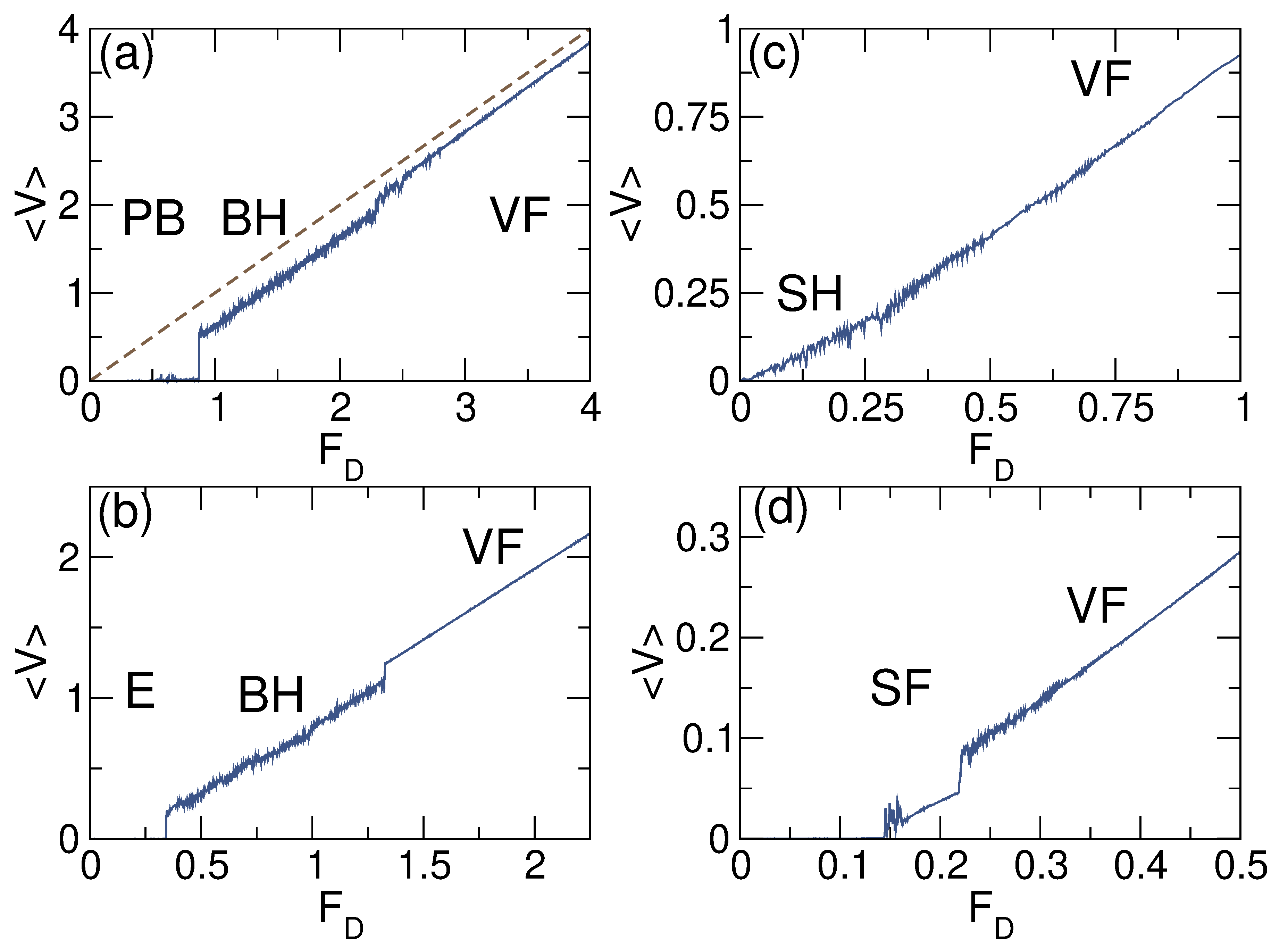}
\caption{$\langle V \rangle$ vs $F_D$ curves in a system with fixed
$\rho=0.32$.
(a) At $\xi = 0.645$ in the bubble regime,
we find an elastic flow phase at small $F_D$,
a plastic bubble flow phase $PB$, a disordered bubble hopping phase $BH$,
and a viscous flow phase $VF$.
The dashed line indicates the expected velocity-force curve
for a free particle.  
(b) At $\xi = 0.6$ in the bubble regime,
we observe elastic flow $E$, bubble hopping $BH$,
and viscous flow $VF$ phases,
but the plastic bubble flow $PB$ phase is absent.
(c) At $\xi = 0.525$ in the stripe regime,
the width of the elastic flow phase reaches a minimum,
and the system passes from a stripe hopping $SH$ state to
viscous flow $VF$ at higher drives.
(d) At $\xi = 0.35$ in the crystal regime,
we find soliton flow $SF$ and viscous flow $VF$.}
\label{fig:4}
\end{figure}

In Fig.~\ref{fig:4}(a), we plot
$\langle V \rangle$ versus $F_D$
for the system from Fig.~\ref{fig:1}
in the bubble regime at a smaller $\xi = 0.645$.
The same four dynamic phases appear that
were found in Fig.~\ref{fig:1},
but the threshold for the elastic-to-plastic bubble transition has dropped
to lower $F_D$,
and the range of the disordered bubble hopping regime is larger and
is accompanied by clear jumps in $\langle V \rangle$ at either end.
At $\xi = 0.6$ in the bubble regime, the $\langle V\rangle$ versus
$F_D$ curve in Fig.~\ref{fig:4}(b) indicates that
the system passes directly from the elastic bubble to
the disordered bubble hopping state,
so that the plastic bubble flow phase is absent.
The transition from the disordered flow regime to the viscous flow
state is even more pronounced, and is accompanied by a marked drop
in the velocity fluctuations.
At $\xi = 0.525$ in Fig.~\ref{fig:4}(c),
the system forms a stripe phase.
There is almost no elastic flow regime,
and we observe only a small region of disordered
stripe-hopping motion followed by a wide viscous flow regime.
In previous work, the velocity-force curves in
the stripe state exhibited
a much stronger transition between the disordered and
viscous regimes \cite{Reichhardt25}, and this is because
the stripes in the previous work were much wider than the
stripes under consideration here.
In the crystal state at $\xi=0.35$,
Fig.~\ref{fig:4}(d) shows that a wide elastic phase
is followed by a soliton-like flow state ($SF$),
where the probe particle moves along the rows and generates
a correlated exchange of particles. This is similar to the
states found in previous work on the crystal state \cite{Reichhardt25}.
There is also a high drive viscous flow regime
where the probe particle moves between rows of
background particles without creating any plastic deformations.
The depinning threshold separating the elastic flow phase from the
plastic or disordered flow phase
is non-monotonic as a function of $\xi$,
and reaches its lowest value in the stripe phase.

\begin{figure}
\includegraphics[width=\columnwidth]{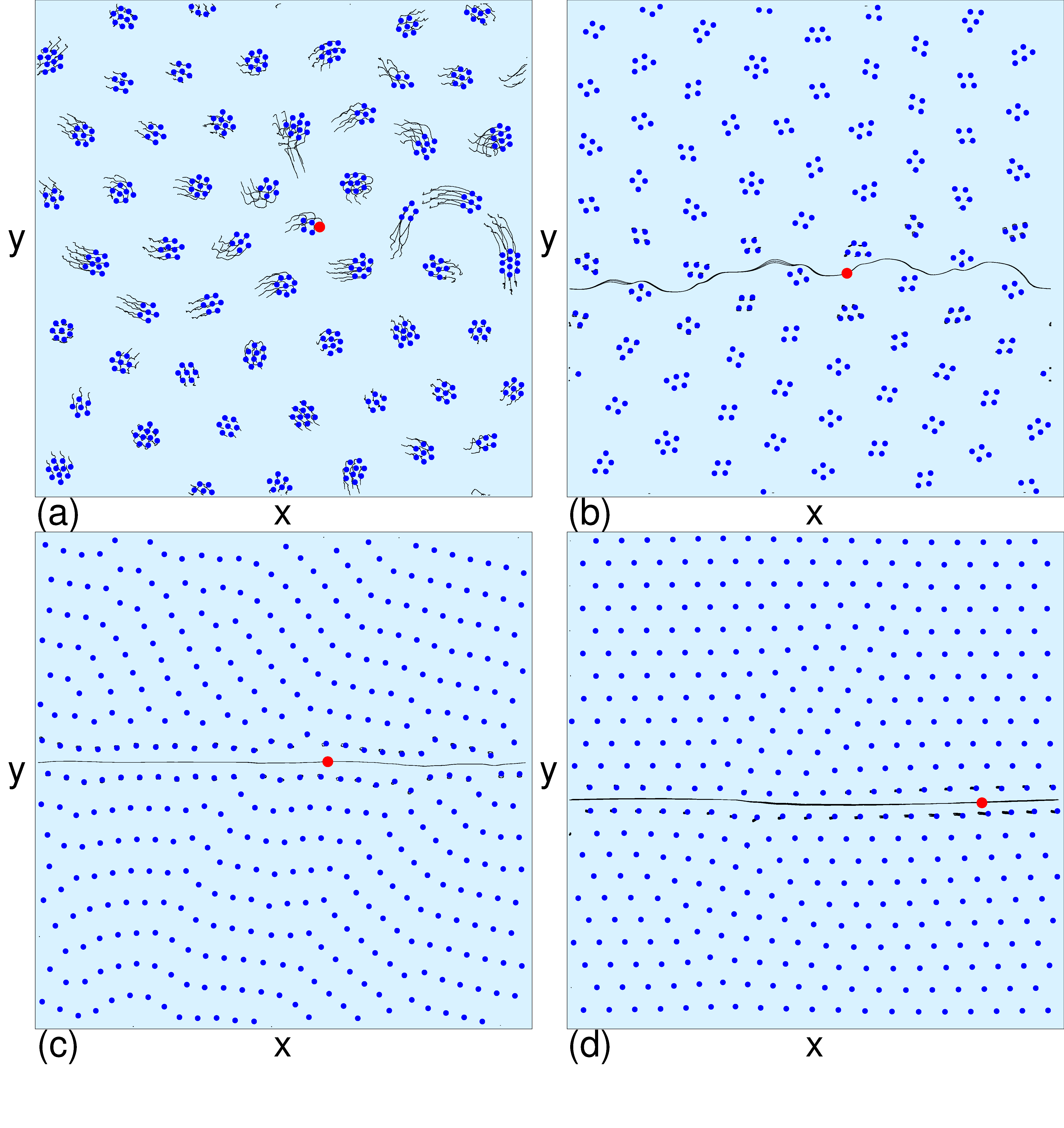}
\caption{
The probe particle (red), background particles (blue), and
particle trajectories (lines) for the system from Fig.~\ref{fig:1}
with
$\rho=0.32$.
(a) Bubble state at $\xi = 0.65$ and $F_D = 0.6$ in the
plastic bubble flow regime.
(b) Bubble state at $\xi = 0.6$ and $F_D = 2.0$
in the viscous flow regime. (c)
Stripe state at $\xi = 0.525$ and $F_D = 1.0$
in a viscous flow regime.
(d) Crystal state at $\xi = 0.475$ and $F_D = 0.7$
in a viscous flow regime.
}
\label{fig:5}
\end{figure}

In Fig.~\ref{fig:5}(a) we show particle positions and trajectories for
the bubble state from Fig.~\ref{fig:4}(a) in the plastic bubble flow regime
at $\xi = 0.65$ and $F_{D} = 0.6$.
Here the clusters are smaller than
the clusters illustrated for the $\xi = 0.7$ system in
Fig.~\ref{fig:2}.
At $\xi=0.6$ and $F_D=2.0$ in
Fig.~\ref{fig:5}(b),
we find a viscous flow regime containing clusters of even smaller size.
At $\xi = 0.525$, a stripe structure appears that is illustrated
in the viscous flow regime for $F_D=1.0$ in
Fig.~\ref{fig:5}(c).
Viscous flow in the crystal state is shown at
$\xi=0.475$ and $F_D=0.7$ in
Fig.~5(d), and we find similar flow as $\xi$ is lowered in the
crystal state.

\begin{figure}
\includegraphics[width=\columnwidth]{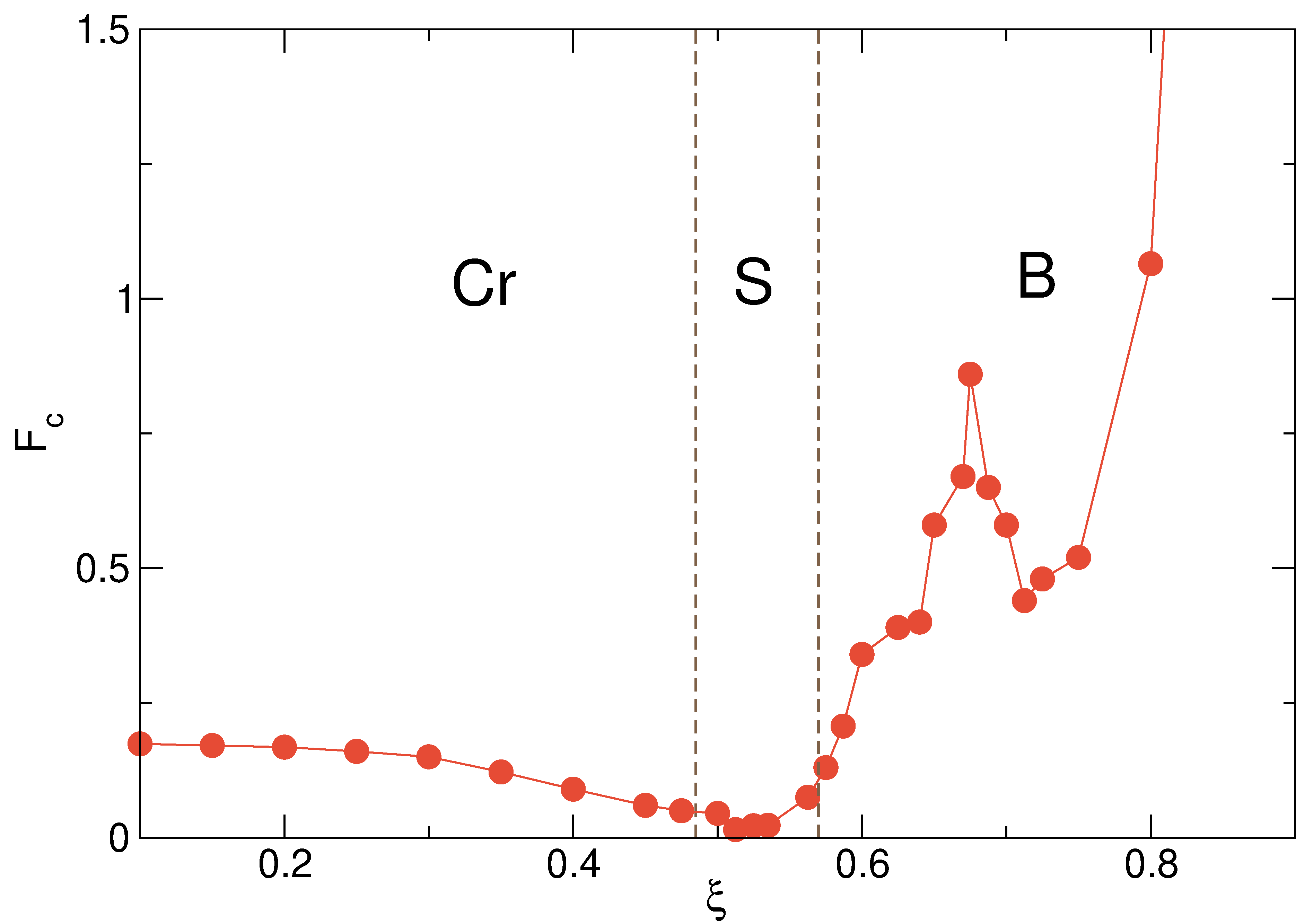}
\caption{The critical depinning force $F_c$ vs
$\xi$ at $\rho = 0.32$
in the crystal (Cr), stripe (S),
and bubble (B) states.
There is a minimum in $F_c$ in the stripe phase.}
\label{fig:6}
\end{figure}

\begin{figure}
\includegraphics[width=\columnwidth]{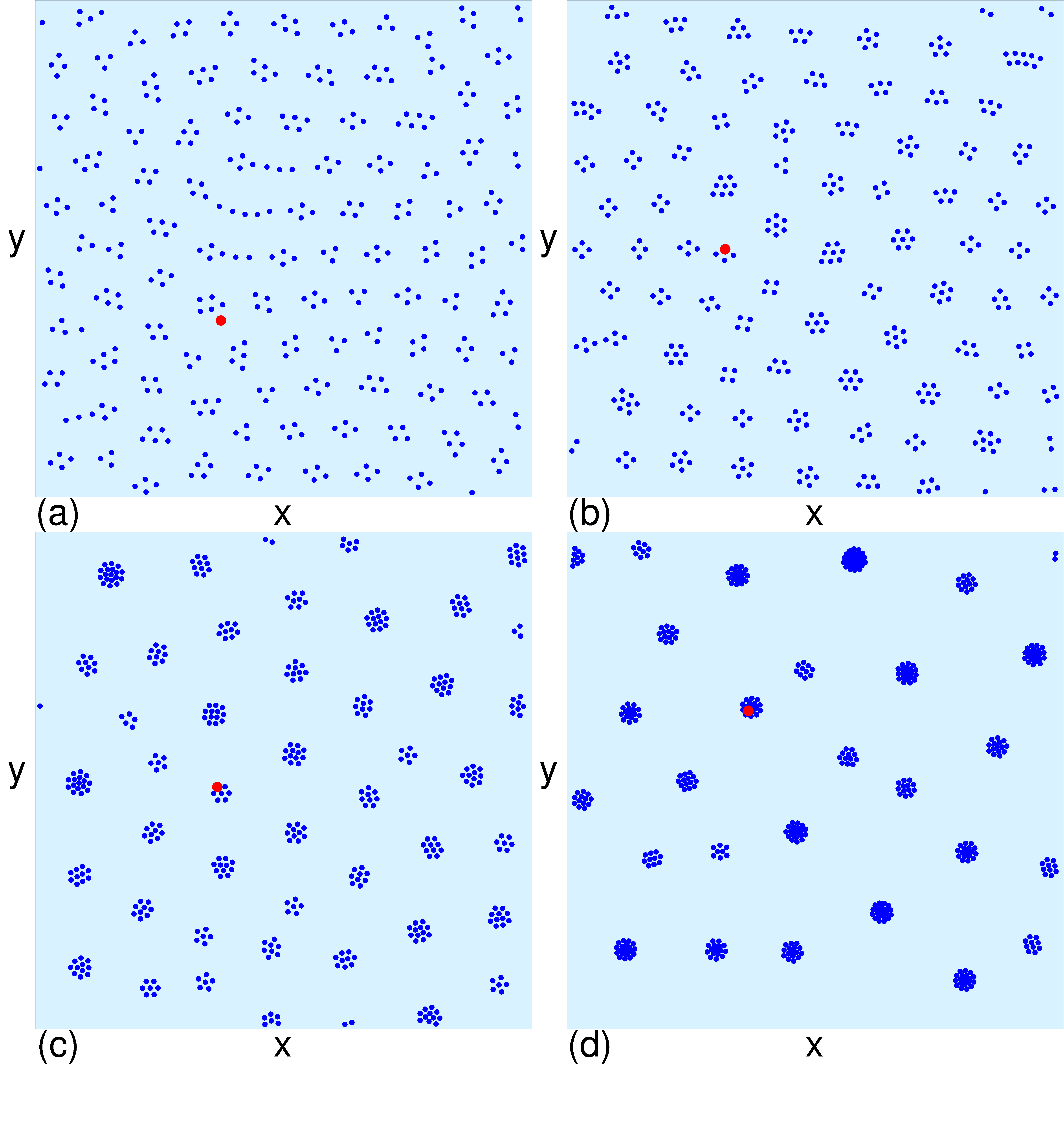}
\caption{Positions of the probe particle (red) and
background particles (blue) for the system in
Fig.~\ref{fig:6} with $\rho=0.32$ at
(a) $\xi = 0.575$, (b) $\xi = 0.625$, (c) $\xi = 0.725$,
and (d) $\xi = 0.85$,
showing that as $\xi$ increases,
the number of bubbles decreases and the number of particles
contained in each bubble increases.
}
\label{fig:7}
\end{figure}

In Fig.~\ref{fig:6} we plot the critical depinning force at which
the elastic phase ends versus
$\xi$ for a system with
$\rho = 0.32$
as it passes through the crystal, stripe, and
bubble states. The elastic phase is defined to end when the probe particle
ceases to drag all of the other particles in the system along with it.
The minimum value of $F_c$ falls in
the stripe phase, and
$F_c$ increases as $\xi$ is decreased into the crystal state or increased
into the bubble state.
Within the bubble phase, $F_c$ is nonmonotonic and passes
through 
a local maximum at $\xi = 0.68$, a local minimum at $\xi=0.71$, 
and then increases for larger $\xi$.
This non-monotonicity is produced by a competition
between the number of bubbles and the size of the bubbles.
For smaller $\xi$ in the bubble phase, the probe particle escapes from
the bubble at $F_c$, but as $\xi$ increases, the bubbles increase
in size and generate a stronger confining force on the probe particle,
so that $F_c$ instead marks the point at which the bubble containing the
probe particle breaks away from the surrounding bubbles and begins to
move through them.
In Fig.~\ref{fig:7}, we show the bubble states for the
system from Fig.~\ref{fig:6} at
$\xi =0.575$, 0.625, 0.725, and $0.85$.
The number of bubbles decreases with increasing $\xi$,
but the number of particles contained inside each bubble increases.
The depinning force for a transition into the plastic bubble flow phase,
where the entire bubble containing the probe particle depins,
is controlled by the effective repulsive interactions between
neighboring bubbles,
and can be approximated as $F_c \propto N_b/R_b^2$,
where $N_b$ is the average number of particles in a bubble
and $R_b$ is the average distance between the bubbles.
With increasing $\xi$, $R_b$ increases since fewer bubbles
are present, which would reduce the depinning force;
however, at the same time, $N_b$ is increasing because each bubble
contains more particles, which would increase the depinning force.
As a result, there is a crossover in the behavior of the depinning
force between $R_b$ dominated for lower $\xi$ and
$N_b$ dominated for larger $\xi$.

\begin{figure}
\includegraphics[width=\columnwidth]{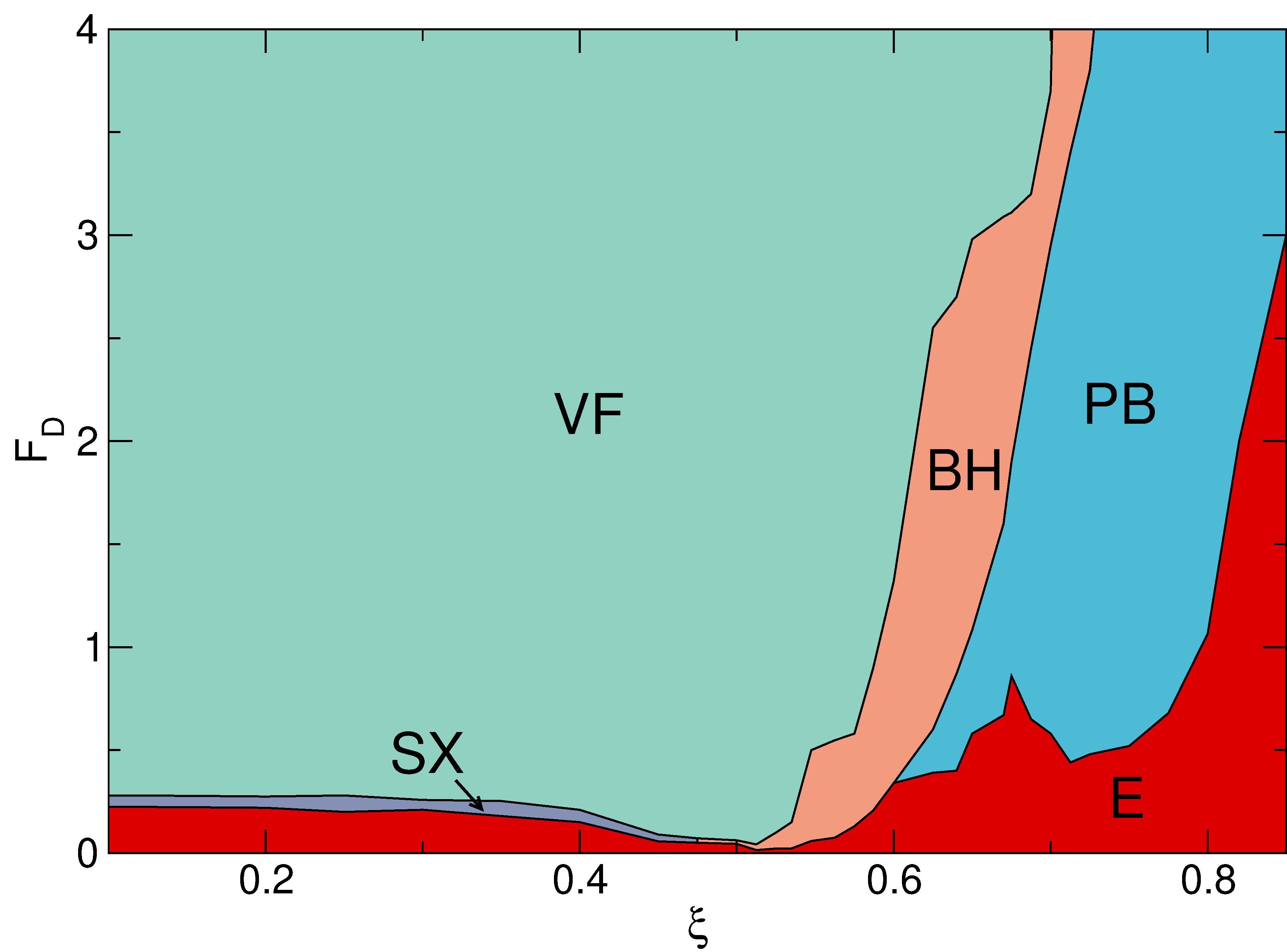}
\caption{
Dynamic phase diagram as a function of $F_D$ vs $\xi$ 
for the system from Fig.~\ref{fig:6} with $\rho=0.32$.
There are
five distinct phases:
an elastic or pinned phase $E$ (red),
a plastic bubble flow phase $PB$ (blue),
a disordered bubble hopping phase $BH$ (orange),
a viscous flow phase $VF$ (green),
and a pulse or soliton phase in the crystal regime $SX$ (purple).
}
\label{fig:8}
\end{figure}

From the different features in the velocity-force curves,
we map out a dynamic phase diagram
as a function of $F_D$ versus $\xi$.
As shown in Fig.~\ref{fig:8}, we identify five phases:
an elastic or pinned phase, a plastic bubble flow phase,
a disordered bubble hopping phase, a viscous flow phase,
and a pulse or soliton phase in the crystal regime.

\section{Varied Density}

\begin{figure}
\includegraphics[width=\columnwidth]{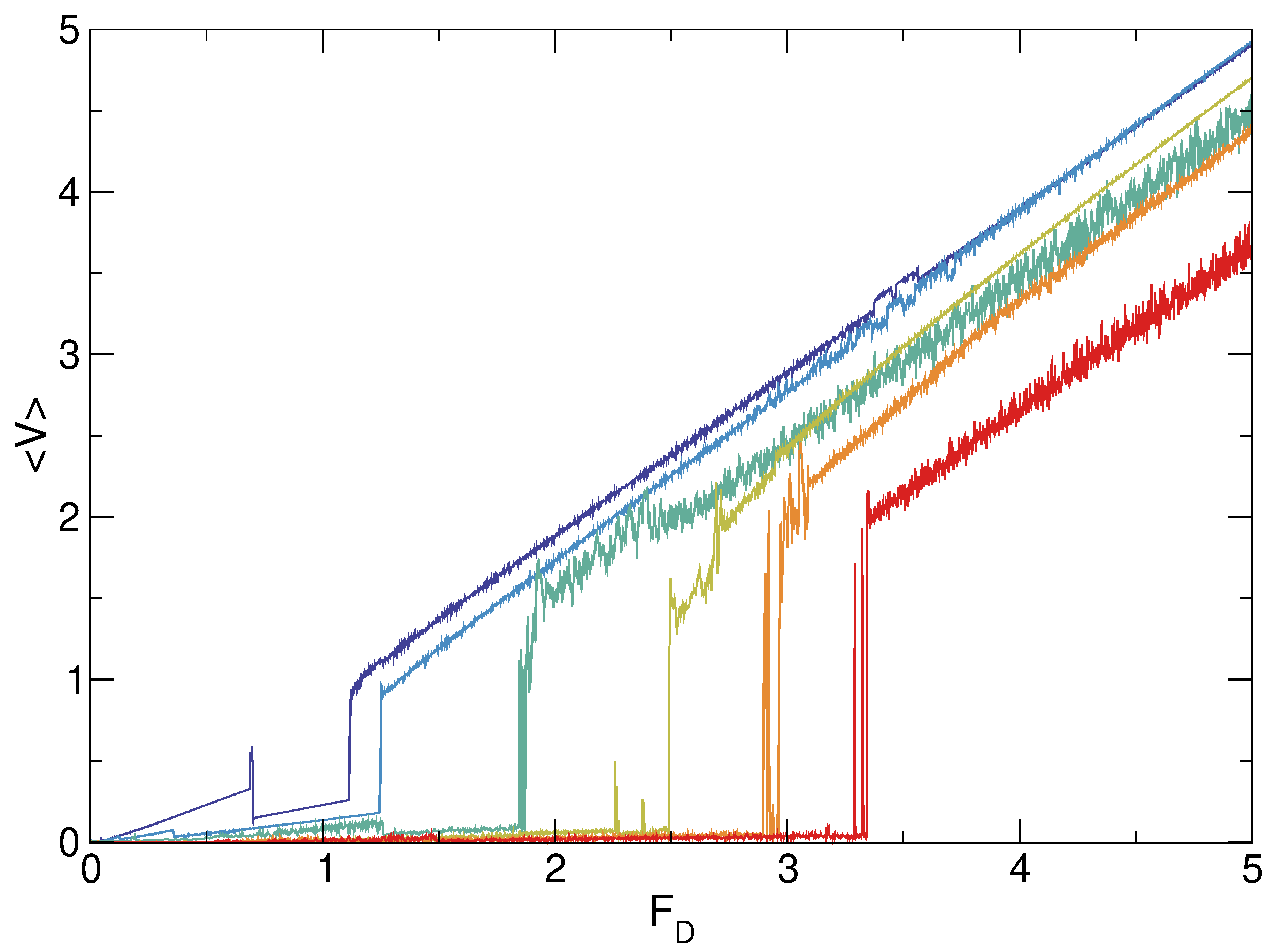}
\caption{The velocity-force curves $\langle V\rangle$ vs $F_D$
for the system from Fig.~\ref{fig:1}
with $\xi = 0.7$ at
$\rho = 0.023$, 0.0617, 0.16, 0.262, 0.32, and $0.37$, from left to right.
}
\label{fig:9}
\end{figure}

We next consider the evolution of the different phases as
we change the particle density $\rho$ while holding
$\xi$ fixed to $\xi=0.7$.
In Fig.~\ref{fig:9}, we show the velocity-force curves for the system
from Fig.~\ref{fig:1} with $\xi = 0.7$ at
varied $\rho = 0.023$, 0.0616, 0.16, 0.262, 0.32, and $0.37$.
For all densities less than $\rho=0.9$,
the system forms bubbles and exhibits a plastic bubble flow phase. 
The transition to the bubble hopping phase shifts
to higher values of $F_D$ as $\rho$ increases,
and there is a critical density $\rho_c$
above which the system transitions from an
elastic to a bubble hopping phase.

\begin{figure}
\includegraphics[width=\columnwidth]{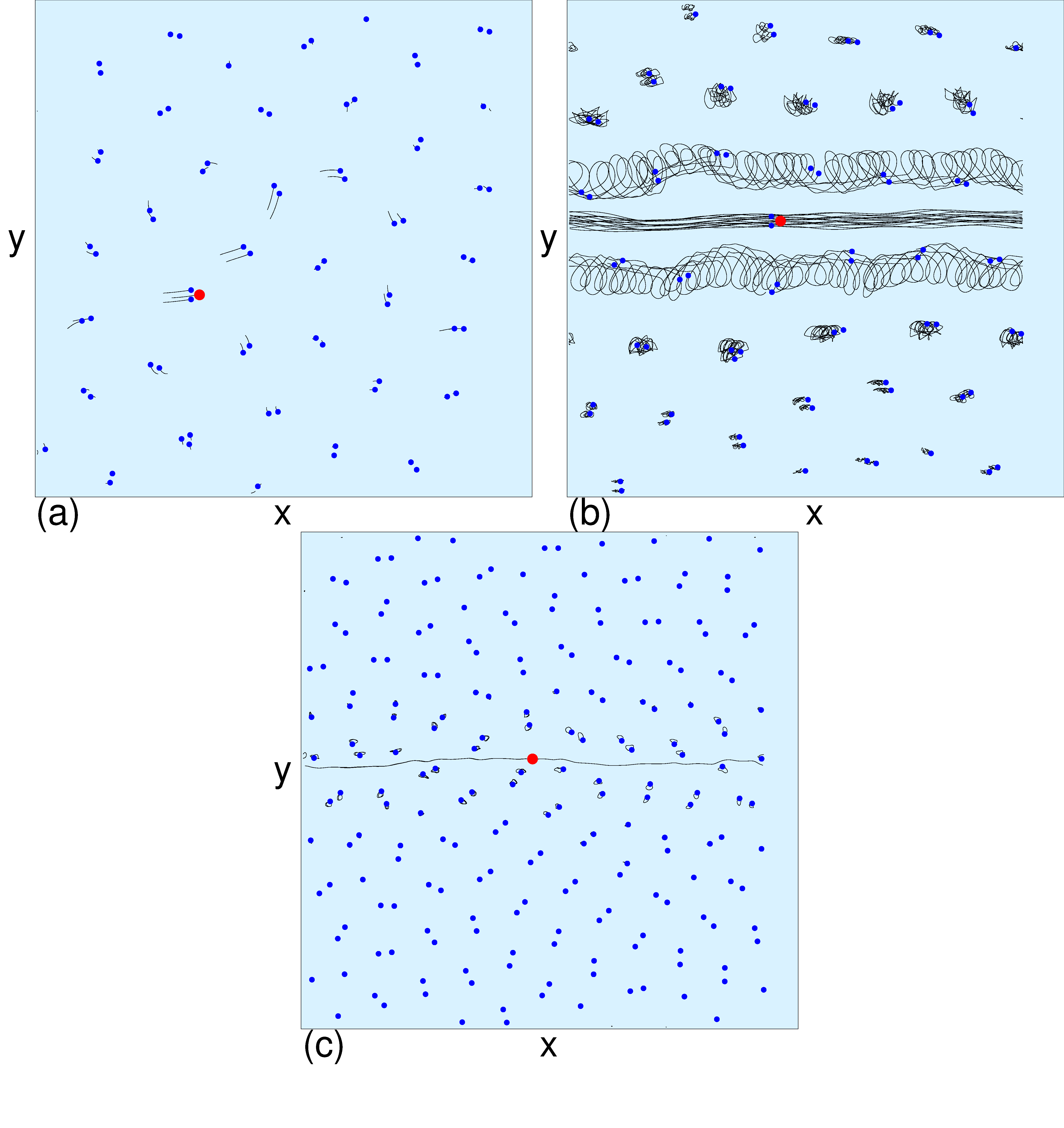}
\caption{The probe particle (red), background particles (blue),
and particle trajectories (lines) for the system in Fig.~\ref{fig:9} with
$\xi=0.7$.
(a) The elastic regime at $\rho = 0.061$ and $F_D = 0.01$,
where the system forms dimer bubbles.
(b) The $\rho=0.061$ system at $F_D = 0.3$ in the plastic bubble flow phase.
(c) The viscous flow phase for the dimer state at
$\rho = 0.16$ and $F_{D} = 1.0$.}
\label{fig:10}
\end{figure}

For $\rho < 0.2$, the bubbles each contain only two particles
that form a dimer.
Figure~\ref{fig:10}(a) shows the trajectories of the probe and background
particles
at $\rho = 0.0617$ and $F_D = 0.01$
in the elastic flow regime.
The probe particle has become attached to a dimer, so the system can be
viewed as a trimer moving through an assembly of dimers.
At this low density, $F_c$ is low and
there is an extended window of plastic bubble motion,
as illustrated in Fig.~\ref{fig:10}(b) at
$\rho=0.061$ and $F_D = 0.3$,
where the moving probe particle and dimer can drag a portion of the background
dimers over a fixed distance.
A portion of the background dimers also undergo correlated rotations
that are the most prominent near the probe particle,
resulting in a
shear banding effect where the dimers in rows that are
adjacent to the probe particle move at a higher velocity
than the dimers in the bulk.
At higher drives, the probe particle breaks out of the trimer state and
begins to jump from dimer to dimer.
The onset of viscous flow in the dimer state occurs at even higher drives,
as shown in Fig.~\ref{fig:10}(c) at
$\rho = 0.16$ and $F_{D} = 1.0$,
where the probe particle passes through the system without
generating any plastic deformations of the background particles.
As the density increases, the background particles form trimers rather than
dimers, but unlike the dimers, the trimers do not rotate in an ordered
fashion and the plastic bubble flow state is more disordered compared
to the dimer case.

\begin{figure}
\includegraphics[width=\columnwidth]{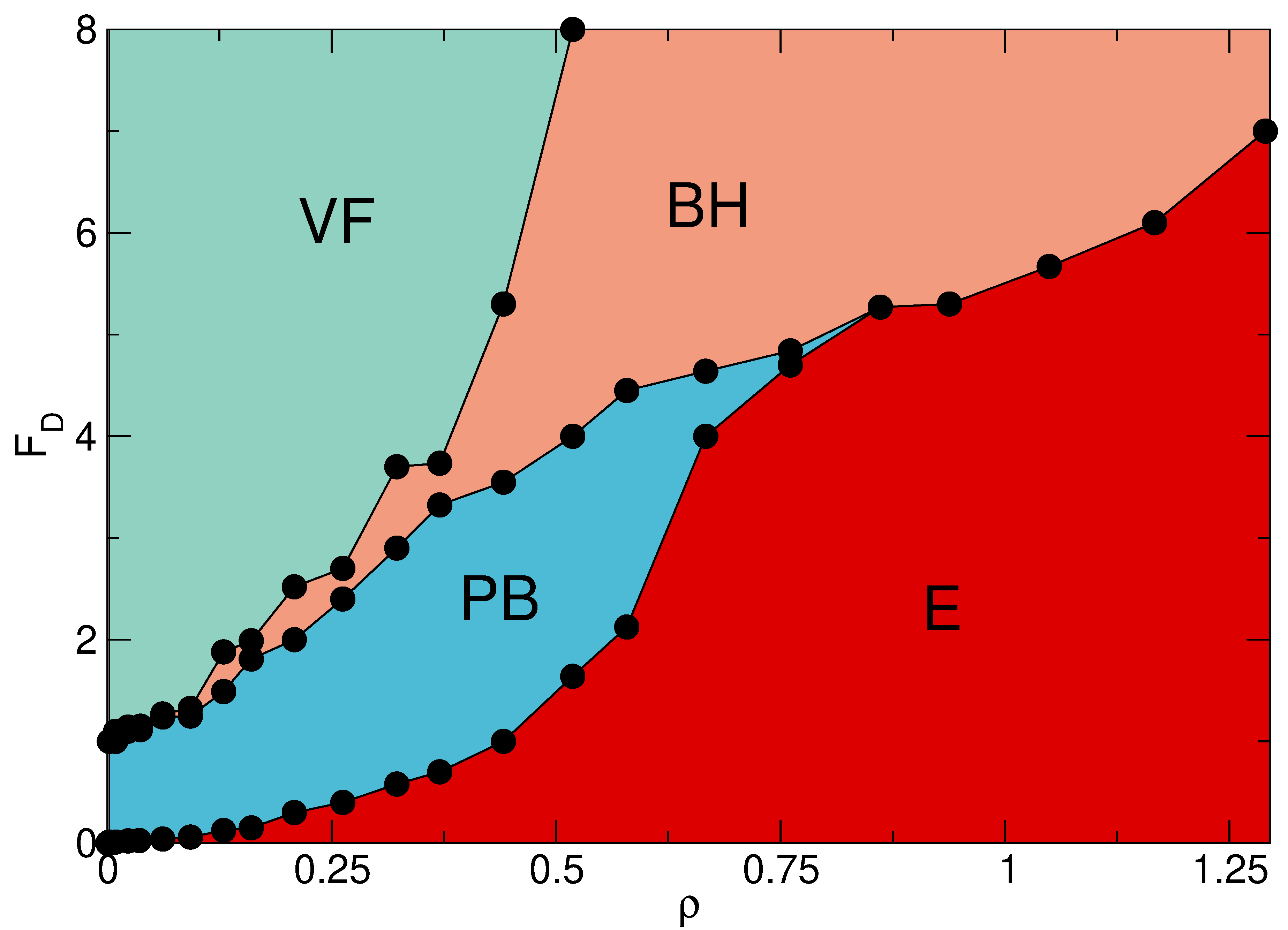}
\caption{Dynamic phase diagram as a function of $F_D$
vs $\rho$ for the system from Figs.~\ref{fig:9} and \ref{fig:10} with
$\xi=0.7$.
We find four distinct phases:
An elastic or pinned phase $E$ (red), 
a plastic bubble flow phase $PB$ (blue), 
a disordered bubble hopping phase $BH$ (orange), 
and a viscous flow phase $VF$ (green). 
There is a critical density $\rho > 0.9$
above which the plastic bubble flow phase does not appear.
}
\label{fig:11}
\end{figure}

In Fig.~\ref{fig:11}, we plot a dynamic phase diagram as a function of $F_D$
versus $\rho$
for the system from Fig.~\ref{fig:9} with $\xi=0.7$ over
a range of parameters containing only bubble states.
The elastic phase monotonically increases in extent as $\rho$ increases,
and a plastic bubble flow phase is present only for $0 < \rho < 0.95$.
For $\rho>0.5$, 
a bubble hopping phase becomes more dominant,
while for smaller $\rho$ there is a 
viscous flow state at large $F_D$.
When $\rho$ is large, each individual bubble contains a large number of
particles, causing the long-range repulsive interaction
between neighboring bubbles to strengthen, and generating
a large barrier for the bubble containing the probe particle
to slide past adjacent bubbles.
In contrast, for small $\rho$, individual bubbles are smaller and the
repulsive interaction between adjacent bubbles is also smaller,
leading to the appearance of 
an extended plastic bubble ($PB$) phase
of the type illustrated for the dimer state
in Fig.~\ref{fig:10}(b).
If $\rho$ is increased beyond the level shown
in Fig.~\ref{fig:11}, eventually a transition occurs from a bubble
state to a stripe state, and in the stripe state,
the critical depinning threshold marking the upper bound of the elastic
flow phase could drop to lower drives
in a manner similar to what has been observed in previous work
\cite{Reichhardt25}.

\begin{figure}
\includegraphics[width=\columnwidth]{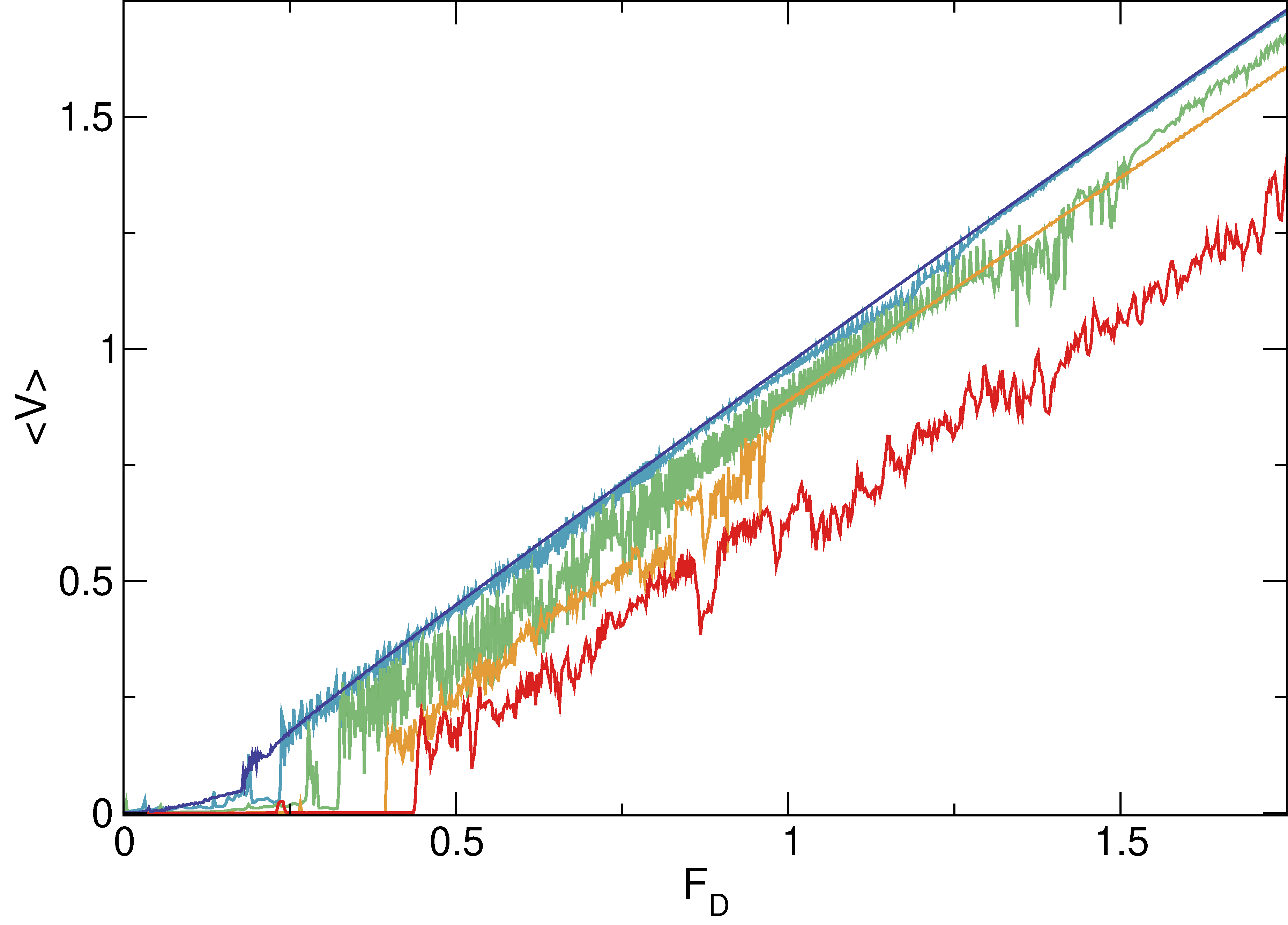}
\caption{Velocity-force curves $\langle V\rangle$ vs $F_D$
for a system with $\xi = 0.6$
at $\rho = 0.0167$, 0.129, 0.208, 0.44, and $0.67$, from left to right.
}
\label{fig:12}
\end{figure}

\begin{figure}
\includegraphics[width=\columnwidth]{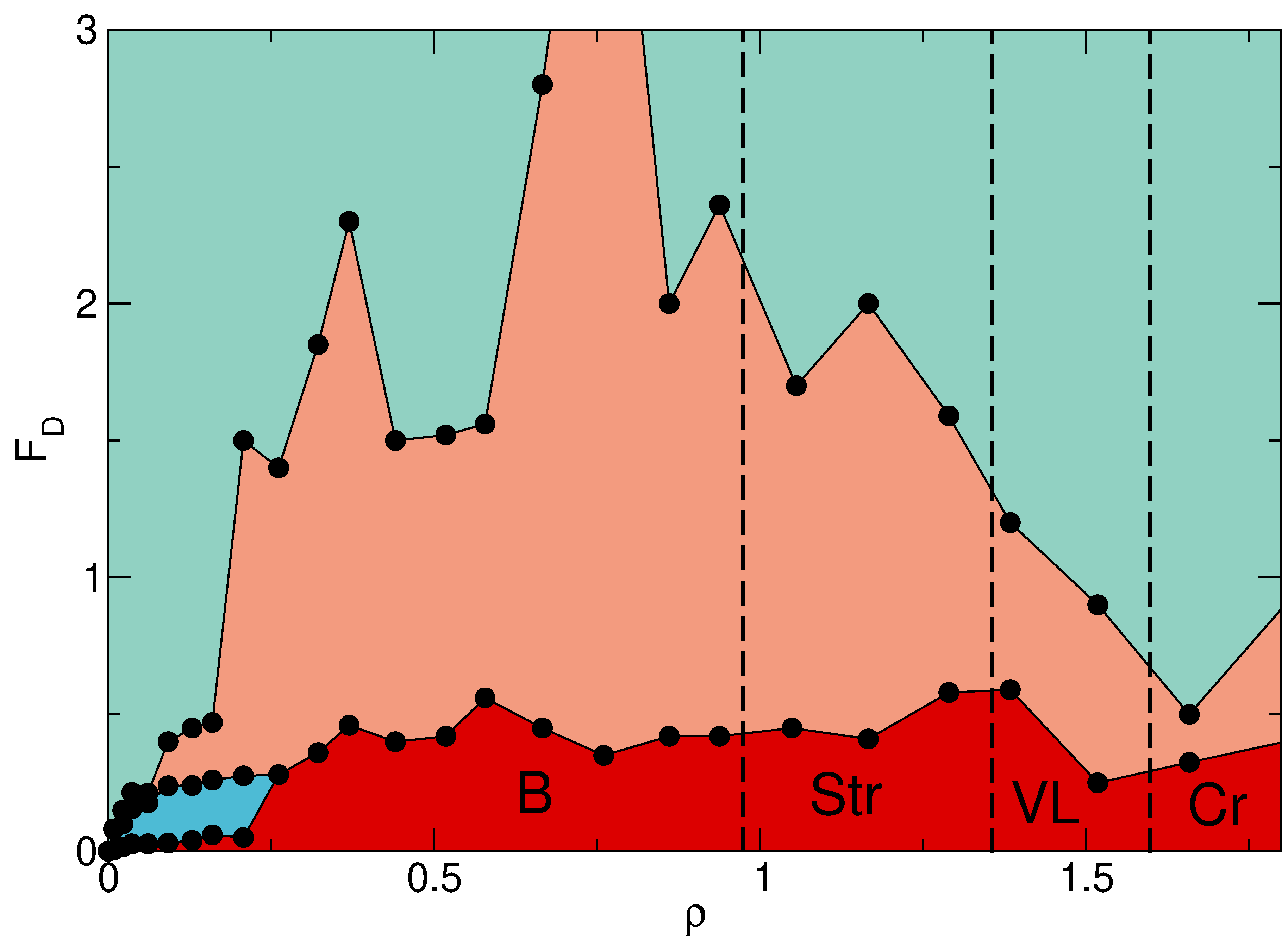}
\caption{Dynamic phase diagram as a function of $F_D$ vs $\rho$
for the system in Fig.~\ref{fig:12} with $\xi = 0.6$.
There are four phases:
an elastic or pinned phase (red),
a plastic bubble flow phase (blue),
a disordered bubble hopping phase (orange),
and a viscous flow phase (green).
The dashed lines highlight the transitions between the
bubble (B), stripe (Str), void lattice (VL), and uniform crystal (Cr) states.
}
\label{fig:13}
\end{figure}

In Fig.~\ref{fig:12}, we plot the velocity-force curves
for a system with $\xi = 0.6$ 
at $\rho = 0.0617$, 0.129, 0.208, 0.44, and $0.67$.
Plastic bubble flow occurs only for $\rho < 0.25$.
A dynamic phase diagram as a function of $F_D$ versus $\rho$ for the
same system appears in Fig.~\ref{fig:13}, were we find
a plastic bubble flow phase at smaller $\rho$
along with elastic, disordered bubble hopping, and viscous flow regimes.
The elastic phase does not grow monotonically
with increasing $\rho$ as was the case for $\xi = 0.7$
in Fig.~\ref{fig:11}, but instead its width
saturates near $\rho = 0.5$.
As $\rho$ increases,
the system forms a stripe phase for $0.87 < \rho < 1.35$,
a void lattice for $1.35 < \rho < 1.81$,
and a dense crystal for $\rho > 1.8$.
The width of the disordered flow phase shows stronger variations
compared to what was observed at a different value of $\xi$ in
previous work \cite{Reichhardt25}. 

\begin{figure}
\includegraphics[width=\columnwidth]{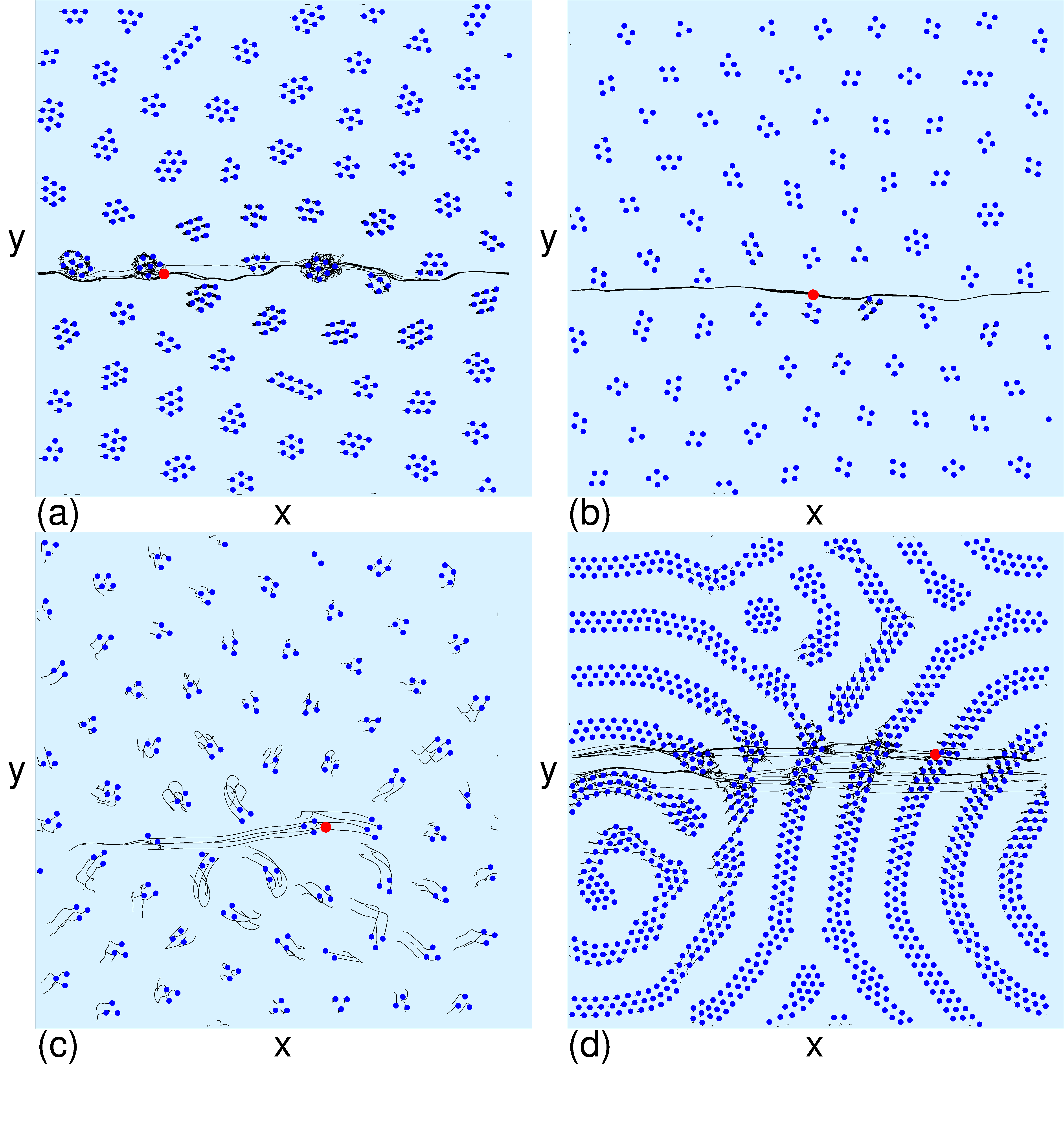}
\caption{The probe particle (red), background particles (blue),
and particle trajectories (lines) for the system in
Fig.~\ref{fig:13} with $\xi=0.6$.
(a) Disordered bubble hopping regime at $\rho = 0.44$ and $F_D = 0.6$.
(b) Viscous flow regime at $\rho=0.32$ and $F_D = 2.0$.
(c) Plastic bubble flow phase at $\rho = 0.16$ and $F_D = 0.1$,
where the system forms trimers.
(d) Disordered hopping state at $\rho=0.9382$ and $F_D = 0.9382$
in a stripe regime.}
\label{fig:14}
\end{figure}

In Fig.~\ref{fig:14}(a), we illustrate the disordered bubble
hopping phase for the system from Fig.~\ref{fig:13} at
$\rho=0.44$ and $F_D = 0.6$.
As the probe particle jumps from bubble to bubble,
it generates plastic deformations or rotations
of the background bubbles, and is able to briefly drag some of the
bubbles.
Figure~\ref{fig:14}(b) shows the viscous flow phase at
$F_D = 2.0$ and $\rho = 0.26$,
while in Fig.~\ref{fig:14}(c) we plot the plastic bubble flow phase
at $F_D=0.1$ and $\rho = 0.16$, where most of the bubble clusters
consist of trimers.
In Fig. 14(d), we show the disordered hopping state for the
stripe regime at $F_D = 0.9382$ and $\rho = 0.9382$, where the
driven particle hops through the stripes.

\begin{figure}
\includegraphics[width=\columnwidth]{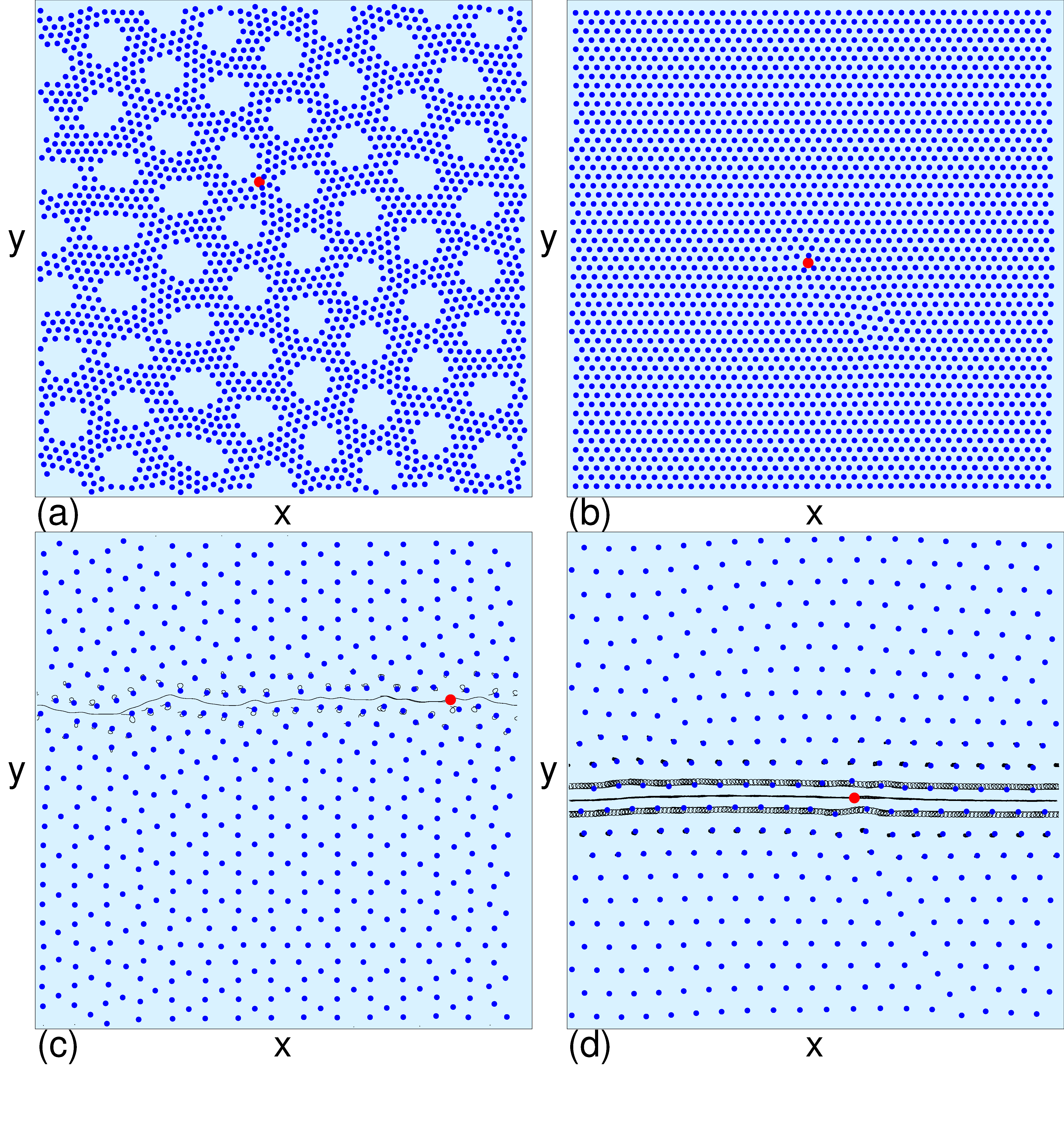}
\caption{(a,b) The probe particle (red) and background particles (blue)
in the elastic or pinned state for the system from Fig.~\ref{fig:13}
with $\xi=0.6$.
(a) Void lattice at $\rho = 1.63$.
(b) Dense crystal phase at $\rho = 1.916$.
(c,d) The probe particle (red), background particles (blue), and
particle trajectories (lines) for $\rho=0.44$ and $F_D=0.3$.
(c) Disordered flow state in a kagome lattice at $\xi=0.525$.
(d) Shear banding in a uniform crystal at $\xi=0.5$.
}
\label{fig:15}
\end{figure}

Images of 
the elastic phase for the system from
Fig.~\ref{fig:13} with $\xi=0.6$
appear in Fig.~\ref{fig:15}(a) at $\rho = 1.63$ where a void lattice
forms and in
Fig.~\ref{fig:15}(b) at $\rho=1.916$ where there is a dense crystal state.
In general,
we see similar phases for other values of $\xi$ and $\rho$; however, there
are some cases for which different orderings
appear at the higher densities.
For example, as shown in Fig.~\ref{fig:15}(c), at
$\xi = 0.525$ and $\rho = 0.44$, a kagome lattice appears and exhibits
disordered flow at $F_D=0.3$.
When $\xi$ is lowered slightly to push the system into the uniform
crystal state, the attractive term can produce
interesting shear banding effects such as that illustrated
in Fig.~\ref{fig:15}(d) at $\xi = 0.5$ and $\rho = 0.44$. Under
the driving force of $F_D=0.3$, 
the probe particle moves between rows of background particles and  can
drag individual background particles a short distance, resulting in local
rotational motion, while the bulk of the background particles
remain immobile.

\section{Discussion}

Our results show that active rheology
in systems with competing interactions can
produce a variety of distinct dynamic behaviors.
In the bubble phase, the probe particle can drag a bubble, causing
the bubble itself to become an effective probe particle
moving through a lattice of other bubbles.
Jumps or changes in velocity-force curve features have
been used previously to distinguish different dynamical phases
in other driven systems with random or periodic substrates, such as 
driven superconducting vortices \cite{Moon96,delaCruz98,Olson98a},
magnetic skyrmions \cite{Reichhardt15aa},
or colloidal particles \cite{Bohlein12},
where there can be pinned phases, soliton motion,
plastic flow, and dynamical ordering at high drives.
Our results indicate that even a single driven probe particle
can also exhibit transitions among different flow phases.
In superconducting vortex systems \cite{Moon96,delaCruz98,Olson98a},
it is known that at higher drives,
the dynamically ordered state forms a coherently moving crystal
or moving smectic. This occurs when the particles are moving
sufficiently rapidly over the substrate that the effectiveness
of the pinning is reduced. The analogous state
for the driven probe particle system
is the viscous flow regime, where the probe particle
is moving fast enough that the surrounding medium
cannot respond to its passage,
while the disordered hopping regimes would be analogous
to the plastic flow state,
where a portion of the particles are pinned while another portion is moving.

Future directions include considering thermal effects,
where thermal creep of the probe particle could occur inside a bubble
and there could also be thermal creep of the entire bubble past the
surrounding bubbles.
It could also be interesting to apply ac rather than dc
driving, as the response of the system would be expected to
depend on the frequency of the driving.
Since the probe particle causes deformations in the surrounding medium,
if multiple driven probe particles were introduced, they might interact
with each other through the background particles, and the effective
interactions
between the probe particles might vary depending upon whether the system
is in the bubble or stripe regime.

\section{Summary} 

We have investigated the driven dynamics of a probe particle in a pattern-forming system with competing long-range repulsion and short-range attraction. We specifically focus on the bubble-forming state and the evolution of the depinning threshold for changing interaction length scales.
This is in contrast to previous work, where the length scale was held fixed
but the ratio of the strength of the attractive and repulsive interaction
terms was varied.
In the bubble state, we find an elastic or pinned regime
where the probe particle is trapped in a bubble
and drags all of the other bubbles along with it.
We also find plastic bubble flow
where the bubble that has trapped the probe particle 
depins and moves past the other bubbles,
causing the bubble itself to act like an effective probe particle.
At higher drives, the probe particle can jump from bubble to bubble
while causing strong plastic deformations of the bubbles,
and at even higher drives, the probe particle moves
sufficiently rapidly that the background particles cannot respond
to it and there are only small rotations of the particles that are
closest to the driven particle.
We show that the depinning threshold is non-monotonic as a
function of changing interaction length scale
and passes through a minimum in the stripe phase.
Even within the bubble state, non-monotonic behavior of the
depinning threshold as a function of increasing interaction length scale
due to a competition 
between a decrease in the number of bubbles present, which lowers the
bubble-bubble interaction strength,
and an increase in the number of particles contained in each
bubble, which increases the bubble-bubble interaction
strength.
We map out a dynamic phase diagram of the different phases
as a function of interaction length and particle density.
We compare our results to the dynamic phases found for
driven particle assemblies, and show that the elastic, plastic,
and viscous flow states of the driven probe particle
are analogous to the pinned, plastic,
and dynamically ordered phases for driven
superconducting vortices, magnetic skyrmions, and colloidal particles.

\begin{acknowledgements}
We gratefully acknowledge the support of the U.S. Department of
Energy through the LANL/LDRD program for this work.
This work was supported by the US Department of Energy through
the Los Alamos National Laboratory.  Los Alamos National Laboratory is
operated by Triad National Security, LLC, for the National Nuclear Security
Administration of the U. S. Department of Energy (Contract No. 892333218NCA000001).
\end{acknowledgements}

\section*{Data Availability Statement}
Data available on reasonable request from the authors.

\nocite{*}
\bibliography{mybib}

\end{document}